\documentclass{vgtc}                          

\ifpdf
  \pdfoutput=1\relax                   
  \usepackage{graphicx}                
  \DeclareGraphicsExtensions{.pdf,.png,.jpg,.jpeg} 
\else
  \usepackage{graphicx}                
  \DeclareGraphicsExtensions{.eps}     
\fi%

\graphicspath{{Figures/}{pictures/}{images/}{./}} 

\usepackage{microtype}                 
\PassOptionsToPackage{warn}{textcomp}  
\usepackage{textcomp}                  
\usepackage{mathptmx}                  
\usepackage{times}                     
\usepackage{cite}                      
\usepackage{tabu}                      
\usepackage{booktabs}                  

\onlineid{1008}

\vgtccategory{Position Paper}

\title{Give Text A Chance:\\
Advocating for Equal Consideration for Language and Visualization}

\author{Chase Stokes\thanks{e-mail: cstokes@ischool.berkeley.edu} %
\and Marti A. Hearst\thanks{e-mail:hearst@berkeley.edu}} %
\affiliation{\scriptsize School of Information \\University of California, Berkeley}

\abstract{
Visualization research tends to de-emphasize consideration of the textual context in which its images are placed. We argue that visualization research should consider textual representations as a primary  alternative to visual options when assessing designs, and when assessing designs,  equal attention should be given to the construction of the language  as to the visualizations.  We also call for a consideration of readability when integrating visualizations with written text. In highlighting these points, visualization research would be elevated in efficacy and demonstrate thorough accounting for  viewers' needs and responses.
} 

\CCScatlist{
  \CCScatTwelve{Human-centered computing}{Visu\-al\-iza\-tion}{}{Visualization design and evaluation methods}
 
}

\begin{document}

\maketitle

\section{Introduction}

For a field called visualization, it is natural that the focus is on examining and implementing visualizations and graphics.  However, it remains important to also factor in the context in which visualizations and graphics appear. When  integrating  visualization with text, the textual component calls for special considerations that we feel are often either overlooked or at least undervalued in assessments of novel interface ideas. 

This occurs despite evidence showing that the title of a visualization tends to influence its interpretation more than the visuals \cite{kong2018frames}, the text is the most memorable part of a recalled infographic \cite{borkin2015beyond},  and that many people prefer text without charts to a combination of charts and text \cite{hearst2019would}. The language within or surrounding a visualization deserves comparable consideration, especially if we consider communication to be the central role of many visualizations.

In this brief essay, we wish to emphasize the following points:

\begin{enumerate}

\item Readability of text should receive high priority when integrating visualizations and graphics with text.   

\item A strong text-only comparison baseline should be considered whenever assessing a  visualization method.
\end{enumerate}

We do not claim that no research addresses these issues, nor that all research in the area falls victim to them. Rather, we want to highlight the issues themselves and encourage these points to be emphasized or explicitly addressed when assessments are done of new visualization techniques and design spaces.  In the remainder of this essay, we elaborate briefly on the reasons we highlight these points.

\section{Text Readability Should Be a Major Concern}

Fluent reading is a major cognitive achievement.  In her seminal book, reading expert Maryanne Wolf summarizes  the impediments to learning to read, the way the brain is transformed to achieve reading fluency, and the complex processes that occur during fluent reading.  She writes \cite{wolf2008proust}:

\begin{quote}
Learning to read changes the visual cortex of the brain.  Because the visual system is capable of object recognition and specialization, the expert reader's visual areas are now populated with cell networks responsible for visual images of letters, letter patterns, and words.  These areas function at tremendous speeds in the expert reader. (pg. 147)
\end{quote}

Why is fluency so important?  Wolf explains this as follows:
\begin{quote}
Fluency does not ensure better comprehension; rather, fluency gives enough extra time to the executive system to direct attention where it is most  needed -- to infer, to understand, to predict, or sometimes to report discordant  understanding and to interpret a meaning afresh. (pg. 131)
\end{quote}

Reading researchers have gathered extensive evidence suggesting that the processing of words occurs in the parafovea before the word is directly fixated on \cite{barach2021emojis}.   Wolf \cite{wolf2008proust} (pg. 148) notes that the preview of what lies ahead makes what follows easier to recognize, contributing further to automaticity. Furthermore, Wolf and other reading researchers \cite{baron2017reading} relay evidence that sustained, deep reading is more easily distracted online, and for a variety of reasons.  Thus, technologists should be mindful of interventions that may contribute to such distractions.

Together, these points should alert visualization researchers to the potential consequences of inserting non-textual graphics within running text.  We should consider: will an inserted image interrupt the highly optimized recognition of text, since it requires a switch from recognizing letters and words to making sense of other kinds of visual objects?

This kind of disruption has been verified in some experimental work; for instance, one study \cite{barach2021emojis} placed  emojis beside  target words with the same or contradictory meanings, finding that the presence of an emoji, relative to its absence, lengthened sentence reading times.  At the same time, total time on the target word decreased in the presence of a semantically matching emoji, which may indicate that congruent emoji can aid understanding of the preceding word.  On the other hand, contradictory emoji slowed reading significantly and increased fixation on the target word.

Images within text can effect other time-sensitive reading tasks, such as document skimming and scanning. Studies show that a common strategy for getting an understanding of a complex document such as a scientific research paper is to look at visual landmarks (figures, tables), and to focus on structural text components (headings) and initial sentences of paragraphs \cite{machulla2018skim}.  
When non-textual visual cues  are interspersed within the text, they are likely to effect this scanning ability in unintended ways.  

For instance, a study of skim reading of hypertext documents that compared pages with and without link navigation found that readers focus on hyperlinks when skimming, and much less on other text and that
readers use hyperlinks as markers for important parts of the text \cite{fitzsimmons2020impact}.  The visual nature of the hyperlinks acted to draw the  attention of the reader, whether this is intended by the author or not.   

Because information visualizations are often larger than letters or words and structurally or visually distinct, their effects on the layout of a document should be considered at the document level as well as the sentence level. However,  studies of designs that integrate visualizations into text often examine individual sentences rather than  effects across a text passage or page. 

It likely would be helpful to examine multimedia -- specifically the combination of visual and textual information.  Multimedia was extensively researched in the 2000's, especially for educational applications.  In highly influential work, Mayer \cite{mayer2002multimedia} provided  rigorous empirical support for nine theory-based effects of how people learn from the combination of words and images. These theories are rooted in a cognitive model founded on dual roles of visual and verbal channels.  One theory he called the ``multimedia effect'', which is the hypotheses that people can learn more deeply from the combination of words and images together than either alone.   The core hypotheses were tested  and confirmed in 11 experiments that contrasted text alone versus text with images.  Mayer also introduced the ``spatial integration effect'' which he and co-authors tested with 5 studies comparing integrated and separated presentation of text and illustrations, finding the integrated presentation lead to better transfer performance than did separated presentation. 

Mayer's findings on the spatial integration effect were confirmed by many later experiments; for instance,  studies have shown that text placed in spatial proximity to explanatory images can reduce cognitive load (e.g., \cite{holsanova2005tracing,holsanova2009reading,zhao2014eye}).
Note that this work did not investigate the embellishment of the text with images \textit{inline}; rather,  text usually was  presented as full sentences or paragraphs (or spoken), and was used to annotate the images or as captions for the images.

The point we wish to emphasize is that visualization researchers should be aware of possible effects on fluent reading, skimming, and other major ways of processing text,  and decide if visualizations inserted within text are intended to affect these.  If so, researchers should discuss what the intended benefits are and if they justify the potential costs on fluent reading.

\section{Studies Should Contain a  Text-Only Baseline}

Visualization work often fails to include a comparison to a no-visualization, text-only version of a design.  This is problematic in light of the evidence that some people prefer to read text or hear spoken language in some circumstances.  This may be a consequence of their familiarity with graphs (which some call visual literacy) or the design of the visualizations themselves \cite{Carpenter1998AMO}.  Either way, this condition should be examined.  

In some of our recent work \cite{hearst2019would}, the motivating research question was how best to show visualizations on small mobile devices in conjunction with a natural language interface.   The text-only condition in this study was not a straightforward choice given the research question, but its inclusion was highly informative.  It showed not only that a significant percentage of participants preferred not to see the charts in a conversational interface, but also a condition under which people were willing to switch from text alone to text plus charts.

The results of Mayer's and other work suggest that the two might work best in combination in at least some educational settings in which visual explanations are likely to be useful, such as showing how a piston works.  That said,  the research community should \textit{prove} that the visual condition as designed is better than the no visual condition.  Having a strong textual baseline as an expectation is likely to compel researchers to think harder about the best way to use text and more closely examine its role. 
This may provide researchers with further information about the usage or impact of the visualization itself as well.

Building a strong text baseline is not necessarily straightforward. The designers of the acclaimed R2D3 machine learning visualization have stated that getting the text right was the most difficult part of the design.\footnote{Personal communication, Tony Chu. http://www.r2d3.us/visual-intro-to-machine-learning-part-1/} 
The importance of and need to focus on high-quality text in association is noted by journalists who write stories that include data and graphics.  Fischer-Baum of the \textit{Washington Post}  noted \cite{fischer}

\begin{quote}
If you do a bad job with your text, people are not going to understand it, no matter how beautiful your visualization is.
\end{quote}

The literature about the best way to formulate titles in visualization is  active, with contrasting views  \cite{Wanzer2021}, and the same could become true of text alone versus text with visualizations, if tested across wide populations. More generally, an expectation of a text-only condition is likely to move the integration of text with visualization forward.

\section{The Way Forward}

Many studies of visualization artifacts are conducted with a small pool of participants, often from computer science departments, and often in the age range of 20-40.   People in these demographics often have better eyesight than older populations, better spatial skills than people in other fields, and better reading fluency than less educated  populations. These factors very likely effect the readability outcomes of these studies.  It is not a new point to state that the study population matters for visualization usability studies; rather we emphasize here that factors that may impact reading of text are often not taken into account in such studies.

We as researchers must ensure that our designs are effective in improving the reader's understanding of the information and thus strongly consider the baseline condition. Experimental conditions should ensure that the addition is a benefit upon the current method of communication, in this case, text. In order to do this, textual conditions should be evaluated with the same rigor as the visualization or graphic itself. 

To fully establish this, user preferences must also be taken into account. As mentioned earlier, some users prefer information via text alone, rather than with the input or addition of charts \cite{hearst2019would}. Preferences should be evaluated as well, to determine the full user experience as well as to examine any impact of preference on comprehension. These additions also assist in iterative design processes and so may be useful to understand not only \textit{if} the visual implementation assists in comprehension but also \textit{where} or \textit{how} the visual implementation is the most helpful to the reader.

\bibliographystyle{abbrv}

\bibliography{bib}

\begin{thebibliography}{10}

\bibitem{barach2021emojis}
E.~Barach, L.~B. Feldman, and H.~Sheridan.
\newblock Are emojis processed like words?: Eye movements reveal the time course of semantic processing for emojified text.
\newblock {\em Psychonomic Bulletin \& Review}, pages 1--14, 2021.

\bibitem{baron2017reading}
N.~S. Baron.
\newblock Reading in a digital age.
\newblock {\em Phi Delta Kappan}, 99(2):15--20, 2017.

\bibitem{borkin2015beyond}
M.~A. Borkin, Z.~Bylinskii, N.~W. Kim, C.~M. Bainbridge, C.~S. Yeh, D.~Borkin, H.~Pfister, and A.~Oliva.
\newblock Beyond memorability: Visualization recognition and recall.
\newblock {\em IEEE transactions on visualization and computer graphics}, 22(1):519--528, 2015.

\bibitem{Carpenter1998AMO}
P.~Carpenter and P.~Shah.
\newblock A model of the perceptual and conceptual processes in graph comprehension.
\newblock {\em Journal of Experimental Psychology: Applied}, 4:75--100, 1998.

\bibitem{fischer}
R.~Fischer-Baum and C.~Estebann.
\newblock Working in a graphics visual storytelling team.
\newblock In {\em Information+ Conference}. 2018.
\newblock https://vimeo.com/301938339.

\bibitem{fitzsimmons2020impact}
G.~Fitzsimmons, L.~T. Jayes, M.~J. Weal, and D.~Drieghe.
\newblock The impact of skim reading and navigation when reading hyperlinks on the web.
\newblock {\em PloS one}, 15(9):e0239134, 2020.

\bibitem{hearst2019would}
M.~Hearst and M.~Tory.
\newblock Would you like a chart with that? incorporating visualizations into conversational interfaces.
\newblock In {\em 2019 IEEE Visualization Conference (VIS)}, pages 1--5. IEEE, 2019.

\bibitem{holsanova2005tracing}
J.~Holsanova, N.~Holmberg, and K.~Holmqvist.
\newblock Tracing integration of text and pictures in newspaper reading.
\newblock {\em Lund University Cognitive Studies}, 125:1--19, 2005.

\bibitem{holsanova2009reading}
J.~Holsanova, N.~Holmberg, and K.~Holmqvist.
\newblock Reading information graphics: The role of spatial contiguity and dual attentional guidance.
\newblock {\em Applied Cognitive Psychology: The Official Journal of the Society for Applied Research in Memory and Cognition}, 23(9):1215--1226, 2009.

\bibitem{kong2018frames}
H.-K. Kong, Z.~Liu, and K.~Karahalios.
\newblock Frames and slants in titles of visualizations on controversial topics.
\newblock In {\em Proceedings of the 2018 CHI Conference on Human Factors in Computing Systems}, pages 1--12, 2018.

\bibitem{machulla2018skim}
T.~Machulla, M.~Avila, P.~Wozniak, D.~Montag, and A.~Schmidt.
\newblock Skim-reading strategies in sighted and visually-impaired individuals: a comparative study.
\newblock In {\em Proceedings of the 11th Pervasive Technologies Related to Assistive Environments Conference}, pages 170--177, 2018.

\bibitem{mayer2002multimedia}
R.~E. Mayer.
\newblock Multimedia learning.
\newblock In {\em Psychology of learning and motivation}, volume~41, pages 85--139. Elsevier, 2002.

\bibitem{Wanzer2021}
D.~L. Wanzer, T.~Azzam, N.~D. Jones, and D.~Skousen.
\newblock The role of titles in enhancing data visualization.
\newblock {\em Evaluation and Program Planning}, 84:101896, 2021.

\bibitem{wolf2008proust}
M.~Wolf.
\newblock {\em Proust and the squid: The story and science of the reading brain}.
\newblock Harper Perennial New York, 2007.

\bibitem{zhao2014eye}
F.~Zhao, W.~Schnotz, I.~Wagner, and R.~Gaschler.
\newblock Eye tracking indicators of reading approaches in text-picture comprehension.
\newblock {\em Frontline Learning Research}, 2(5):46--66, 2014.

\end{thebibliography}

\end{document}